\documentclass{agujournal2019}
\usepackage{soul}
\usepackage{tabularx}
\usepackage{booktabs}
\draftfalse

\journalname{Transport in Porous Media}

\begin{document}

%% ------------------------------------------------------------------------ %%
%  Title
%
% (A title should be specific, informative, and brief. Use
% abbreviations only if they are defined in the abstract. Titles that
% start with general keywords then specific terms are optimized in
% searches)
%
%% ------------------------------------------------------------------------ %%

% Example: \title{This is a test title}

\title{Python workflow for segmenting multiphase flow in porous rocks}

%% ------------------------------------------------------------------------ %%
%
%  AUTHORS AND AFFILIATIONS
%
%% ------------------------------------------------------------------------ %%

% Authors are individuals who have significantly contributed to the
% research and preparation of the article. Group authors are allowed, if
% each author in the group is separately identified in an appendix.)

% List authors by first name or initial followed by last name and
% separated by commas. Use \affil{} to number affiliations, and
% \thanks{} for author notes.
% Additional author notes should be indicated with \thanks{} (for
% example, for current addresses).

% Example: \authors{A. B. Author\affil{1}\thanks{Current address, Antartica}, B. C. Author\affil{2,3}, and D. E.
% Author\affil{3,4}\thanks{Also funded by Monsanto.}}

\authors{Catherine Spurin\affil{1}, Sharon Ellman\affil{2}, Dane Sherburn\affil{3}, Tom Bultreys\affil{2}, and Hamdi A. Tchelepi\affil{1}}
%\affiliation{Department of Energy Resources Engineering, Stanford University, %USA}

\affiliation{1}{Department of Energy Science \& Engineering, Stanford University}
\affiliation{2}{Department of Geology, Ghent}
\affiliation{3}{Work done as a Machine Learning Alignment Theory Scholar at the Stanford Existential Risks Initiative}
% \affiliation{2}{Second Affiliation}
% \affiliation{3}{Third Affiliation}
% \affiliation{4}{Fourth Affiliation}

%(repeat as many times as is necessary)

%% Corresponding Author:
% Corresponding author mailing address and e-mail address:

% (include name and email addresses of the corresponding author.  More
% than one corresponding author is allowed in this LaTeX file and for
% publication; but only one corresponding author is allowed in our
% editorial system.)

% Example: \correspondingauthor{First and Last Name}{email@address.edu}

\correspondingauthor{Catherine Spurin}{cspurin@stanford.edu}

%% Keypoints, final entry on title page.

%  List up to three key points (at least one is required)
%  Key Points summarize the main points and conclusions of the article
%  Each must be 140 characters or fewer with no special characters or punctuation and must be complete sentences

% Example:
% \begin{keypoints}
% \item	List up to three key points (at least one is required)
% \item	Key Points summarize the main points and conclusions of the article
% \item	Each must be 140 characters or fewer with no special characters or punctuation and must be complete sentences
% \end{keypoints}

%% ------------------------------------------------------------------------ %%
%
%  ABSTRACT and PLAIN LANGUAGE SUMMARY
%
% A good Abstract will begin with a short description of the problem
% being addressed, briefly describe the new data or analyses, then
% briefly states the main conclusion(s) and how they are supported and
% uncertainties.

% The Plain Language Summary should be written for a broad audience,
% including journalists and the science-interested public, that will not have 
% a background in your field.
%
% A Plain Language Summary is required in GRL, JGR: Planets, JGR: Biogeosciences,
% JGR: Oceans, G-Cubed, Reviews of Geophysics, and JAMES.
% see http://sharingscience.agu.org/creating-plain-language-summary/)
%
%% ------------------------------------------------------------------------ %%

\vspace{1cm}

This paper has been peer reviewed and  published in Transport in Porous Media.

\url{https://doi.org/10.1007/s11242-024-02136-2}
\vspace{1cm}

%% \begin{abstract} starts the second page
\vspace{1cm}
\begin{center}
\textbf{Keywords}
\begin{itemize}
    \item Multiphase flow 
    \item Porous media 
    \item Segmentation 
    \item Image processing 
    \item Sensitivity analysis
\end{itemize}

%\textbf{Related research article} 

%\begin{table}[htbp]
%\caption{Specification Table}
%\begin{tabularx}{\textwidth}{lX}
%\toprule
%\textbf{Subject area} & Environmental Science, Hydrogeology, Image processing %\\
%\addlinespace % Add extra space between rows
%\textbf{More specific subject area} & Multiphase flow in porous media \\
%\addlinespace % Add extra space between rows
%\textbf{Method name} & Image segmentation \\
%\addlinespace % Add extra space between rows
%\textbf{Name and reference of original method} & Automatization of workflow used in \cite{spurin2020real}. Now implemented in python. \\
%\addlinespace % Add extra space between rows
%\textbf{Resource availability} & The datasets and code base are available %from \url{https://github.com/cspurin/image_processing} \\
%\bottomrule
%\end{tabularx}
%\end{table}
\end{center}

\begin{abstract}
X-ray micro-computed tomography (X-ray micro-CT) is widely employed to investigate flow phenomena in porous media, providing a powerful alternative to core-scale experiments for estimating traditional petrophysical properties such as porosity, single-phase permeability or fluid connectivity. However, the segmentation process, critical for deriving these properties from greyscale images, varies significantly between studies due to the absence of a standardized workflow or any ground truth data. This introduces challenges in comparing results across different studies, especially for properties sensitive to segmentation. To address this, we present a fully open-source, automated workflow for the segmentation of a Bentheimer sandstone filled with nitrogen and brine. The workflow incorporates a traditional image processing pipeline, including non-local means filtering, image registration, watershed segmentation of grains, and a combination of differential imaging and thresholding for segmentation of the fluid phases. Our workflow enhances reproducibility by enabling other research groups to easily replicate and validate findings, fostering consistency in petrophysical property estimation. Moreover, its modular structure facilitates integration into modeling frameworks, allowing for forward-backward communication and parameter sensitivity analyses. We apply the workflow to exploring the sensitivity of the non-wetting phase volume, surface area, and connectivity to image processing. This adaptable tool paves the way for future advancements in X-ray micro-CT analysis of porous media.

\end{abstract}

%%%%% MAIN TEXT 
\section{Introduction}
Recent advances in X-ray tomography has revolutionized the field of porous media research, by enabling the \textit{in-situ} visualization of multiple fluid phases within the pore space of rocks \cite{wildenschild2013x, blunt2017multiphase}. The underlying physics governing fluid flow has been explored using synchrotron tomography \cite{scanziani2020situ, berg2013real, spurin2020real}, and fast lab-based micro computed tomography (Micro-CT) \cite{bultreys2016fast, mascini2021fluid, bultreys2022x}. These findings are important in subsurface applications such as CO$_2$ storage \cite{garing2017pore, spurin2024role, herring2013effect} and hydrogen storage \cite{zhang2023pore, jangda2023pore, thaysen2023pore}. 

Consequently, large amounts of 4D data has been generated that has to be segmented into the various fluid phases occupying the pore space, and the rock grains themselves. Segmentation is essential for quantitatively analyzing flow in porous media. However, segmentation can be a difficult and time-consuming process. The key issues are: (1) the lack of ground truth data for benchmarking, (2) image artefacts, (3) features below or near the image resolution, and (4) difficulty in deciding a metric for segmentation. These issues can be addressed by one, or a combination of, (1) filtering the images \cite{sheppard2004techniques, kaestner2008imaging}, (2) only analyzing features above image resolution \cite{spurin2020real, huang2021effect}, (3) alternate segmentation methods, such as machine learning, that include features such as texture \cite{arganda2017trainable, purswani2020evaluation, garfi2020sensitivity} or super resolution \cite{zhou2022neural} while conventional segmentation relies on greyscale values. Due to the lack of ground truth data, results are often assessed qualitatively, highlighting the need for open access segmentation methods \cite{garfi2020sensitivity, sheppard2014techniques, sheppard2004techniques}.

A range of image processing workflows exists, with many researchers opting for commercial software such as Avizo \cite{andrew2015imaging, berg2014multiphase, gao2017x}. The identification of boundaries has been observed to be sensitive to the image processing workflow \cite{garfi2020sensitivity}, but the impact of this is not typically considered when comparisons are made between the final segmented results of different groups. This is due to the absence of a fully-integrated, fast workflow. 

In this work, we present an accurate, consistent and fast image processing workflow that is fully open-source and can be integrated into additional python workflows. We explore the sensitivity of the non-wetting phase volume, surface area, and connectivity measurements to the image processing workflow. With this, we are able to quantify the degree of uncertainty in these measurements introduced by the image processing workflow used.

\section{Data and Methods}

\subsection{Dataset}

In this study we explore the sensitivity of multiphase flow properties in a cylindrical Bentheimer sandstone sample, 25~mm in diameter and 45~mm in length. The experiments are described in detail in previous work \cite{spurin2024role}. The voxel size for the images acquired was 20~$\mu$m. The analysis was performed on a subvolume of $200^3$~voxels, i.e. 16~mm$^3$. The fluids used were brine (deionized water doped with 17 wt\% KI to improve the X-ray contrast) and nitrogen gas. The system was pressurized to 5 MPa to minimize the compressibility of nitrogen, with an additional 2 MPa of confining pressure.

\subsection{Python Workflow}

\begin{figure}[!h]
	\centering
\includegraphics[width=0.9\textwidth]{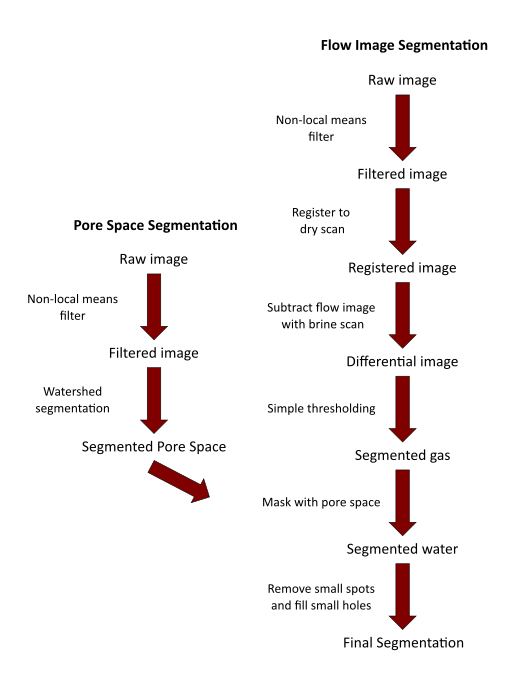} 
	\caption{Image processing workflow for segmentation. The pore space segmentation requires a high quality image of the sample without brine present, as brine is highly attenuating. The segmentation of the flow images requires the segmented pore space, the image of the sample with both fluids present, and an image with just the wetting phase (typically brine) present.}
	\label{workflow}
\end{figure}

In this work, we describe: (1) the segmentation of the pore space, and (2) the segmentation of the fluids within the pore space, using our open-source python workflow. The workflow for the segmentation of the pore space is discussed in Section \ref{pore_segmentation}, and the segmentation of the fluids is discussed in Section \ref{fluid_segmentation}. A schematic of the full workflow is shown in Figure \ref{workflow}. 

The pore space segmentation workflow requires a high quality scan of the sample i.e. no movement, beam hardening or ring artefacts post reconstruction. A lot of these features are removed in the reconstruction, if the reconstruction parameters are selected correctly, \cite{kaestner2008imaging, munch2009stripe}. Thus they are not discussed in this work. The high quality scan can be acquired with either air or deionized water occupying the pore space (but not brine because it has a higher X-ray attenuation). Using deionized water pressurized to experimental conditions minimizes the probability of grain movement between this scan and subsequent scans where fluids are present in the pore space. However, for samples with high clay content, deionized water encourages the swelling of clay minerals \cite{aksu2015swelling,zhang2021impact}.

The flow image segmentation workflow relies on subtracting the image with both fluids present, with an image where the sample is fully saturated with the wetting phase (referred to as the brine scan in Figure \ref{workflow}, as brine is typically the wetting phase). This is the easiest way to automate the segmentation process, as the subtraction leads to one peak in the greyscale histogram of the pore space (which corresponds to the location of the non-wetting phase). It also makes the segmentation more robust in instances where the contrast between the fluids present is less extreme. The subtraction step can be removed if the contrast of the fluids is sufficient, and may be necessary in situations where a scan of the sample fully saturated with the wetting phase was not possible.

\subsection{Segmentation of the Pore Space}
\label{pore_segmentation}
For the segmentation of the pore space, there are two steps to the process: (1) filtering the image using a non-local means filter, and (2) segmentation of the pore space and grains using the watershed segmentation method. 

The non-local means filter reduces noise, while maintaining boundaries between fluid phases \cite{buades2011non}. This is done by selecting a small patch of voxels, and comparing the voxels of interest to their surrounding neighbours; if the pixels have similar greyscale values, they are averaged, thus reducing noise. The aim is to have the maximum amount of filtering while maintaining boundaries \cite{berg2018generation}.

\begin{figure}[!h]
	\centering
\includegraphics[width=0.9\textwidth]{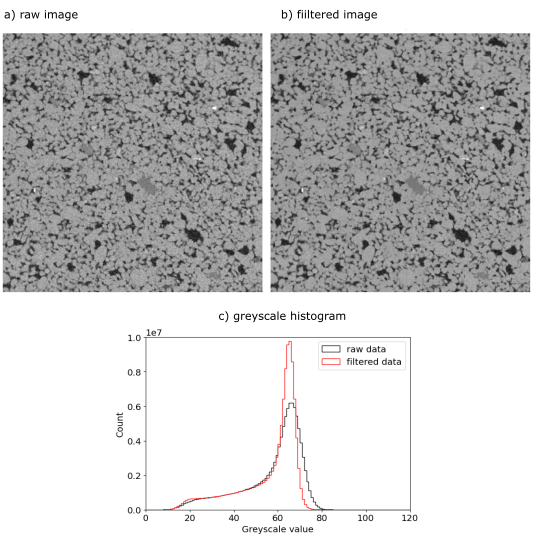} 
	\caption{The impact of non-local means filtering on images: a) raw image, b) filtered image, and c) histogram of greyscale values for the raw and filtered images.}
	\label{filter_dry_scan}
\end{figure}

 \begin{figure}[!h]
	\centering
\includegraphics[width=0.9\textwidth]{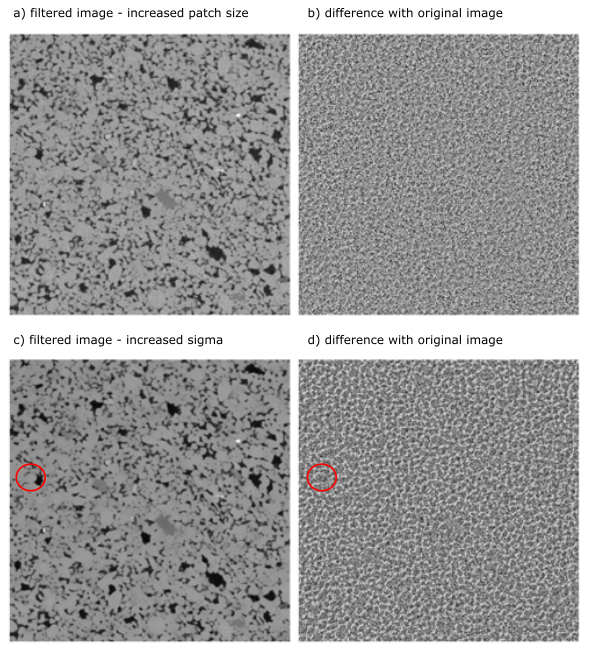} 
	\caption{a) filtered image with an increased patch size, b) difference image of the filtered image in a) with the original image in Figure \ref{filter_dry_scan}a, c) filtered image with an increase noise standard deviation (sigma), and d) difference image of the filtered image in c) with the original image in Figure \ref{filter_dry_scan}a.}
	\label{filter_different_params}
\end{figure}

The parameters that can be varied are: (1) the mode of filtering; either uniform spatial weighting on the patches or a spatial Gaussian weighting,
(2) the noise standard deviation, which can be estimated automatically, 
(3) the patch size, and
(4) the maximal distance in voxels where to search for the patches used for denoising, or search area. The patch size, and search area heavily control the run time of the filtering function. 

The best filtering for this data set was chosen as uniform spatial weighting, the noise standard deviation set to 3 times the automatically estimated value, a patch size of 3 voxels, and a search area of 6 voxels. This is shown in Figure \ref{filter_dry_scan}. By parallel processing (splitting the array into overlapping chunks and running the filtering in parallel on the chunks), the image filtering time for an example $1400 \times 1400 \times 600$ voxel image reduces from approximately 1 hour to 10 minutes on a machine with 32GB RAM, with no reduction in filter quality. The same function took over 4 hrs using the commercial software Avizo. Doubling the patch size has little  observable impact on image quality (see Figure \ref{filter_different_params}a) but reduces the porosity by 5\% and increases the run time of the filtering. Increasing the noise standard deviation (sigma) to 5 times the estimated value makes the images look smoother (see Figure \ref{filter_different_params}c), but the grain boundaries begin to be influenced by the filtering process \cite{buades2005non, buades2011non}. An example is highlighted in the red circle in Figure \ref{filter_different_params}c and d. 

The pore space of the rock sample is typically darker than the surrounding rock grains. However, there is overlap in the lightest regions of the pore space, and the darkest regions of the rock grains, leading to some overlap in greyscale values (see Figure \ref{filter_dry_scan}c). As a result, a simple thresholding cannot accurately capture the pore space. Instead we use a watershed segmentation that looks at greyscale values, and gradients within the greyscale distribution. 

The watershed algorithm is a popular image segmentation technique that treats the grayscale image as a topographic surface, where voxel intensities represent elevation \cite{kaestner2008imaging, brun2010pore3d}. Grayscale gradients are used to identify the edges of objects, with higher gradients indicating steeper slopes. The algorithm simulates water flooding from the lowest elevations, causing it to fill catchment basins (regions) until it reaches watershed lines (boundaries), effectively segmenting the image based on these gradients \cite{bleau2000watershed, kornilov2022review}.

In our workflow, the only user input for the watershed segmentation is the value, below which, only pore space exists. The value, above which, only grains exist is calculated from the histogram in Figure \ref{filter_dry_scan}c. This is because, for most applications in subsurface porous media, the porosity is less than 30\% of the total image, and so while a peak does not always exist in the greyscale histogram for the pore space, there is a peak associated to the rock grains, which make up the majority of the voxels in a given image. A Gaussian distribution is fitted through this to get the threshold value for the grains. 

\begin{figure}[!htbp]
	\centering
\includegraphics[width=1\textwidth]{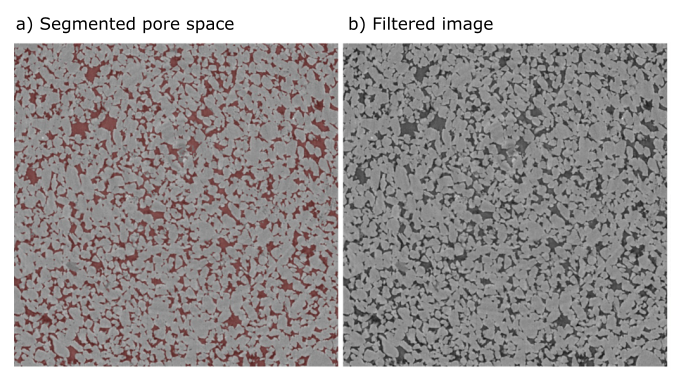} 
	\caption{a) the segmented pore space overlain on the filtered image, and b) the filtered image. }
	\label{watershed_segmentation}
\end{figure}

Small objects are removed at the end of the segmentation to improve the signal to noise ratio. These features are too small to be accurately analyzed, as they are close to the image resolution. The final segmentation of the pore space is shown in Figure \ref{watershed_segmentation}a. 

\subsection{Segmentation of Flow Images}
\label{fluid_segmentation}
\begin{figure}[!htbp]
	\centering
\includegraphics[width=1.0\textwidth]{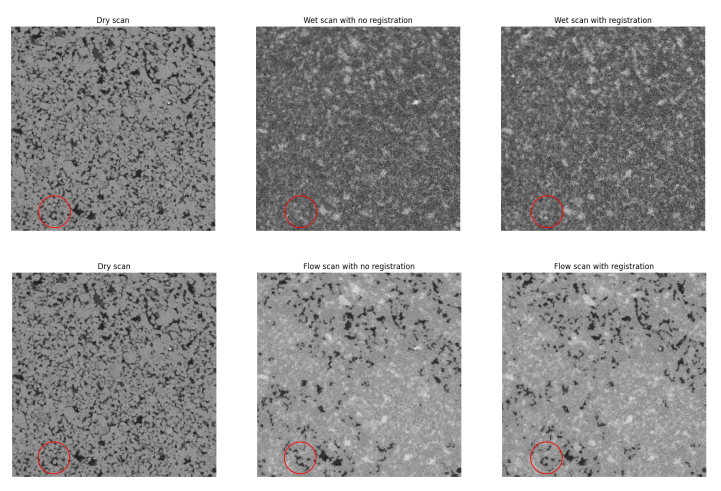} 
	\caption{The impact of registering the wet scan (image with just the wetting phase present) and the flow scan (image with both phases present) with the high quality dry scan.}
	\label{registration}
\end{figure}

The flow images are registered to the initial dry image used for the segmentation of the pore space. This is necessary because small movements can occur during scanning. Without correcting for these, the segmentation of the fluids will be unsuccessful. The registration process is highlighted in Figure \ref{registration}. Without registering, the masking of the flow image with the segmented pore space leads to erroneous results, an example of this is highlighted by the red circle in Figure \ref{registration}. 

\begin{figure}[!htbp]
	\centering
\includegraphics[width=1.0\textwidth]{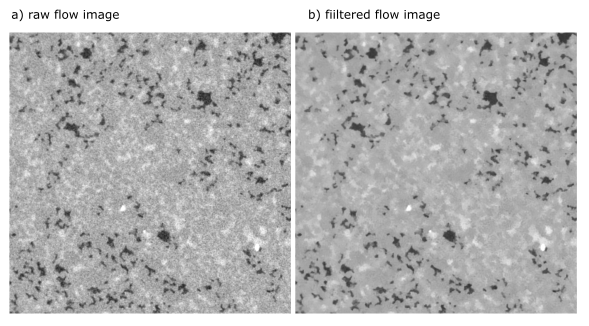} 
	\caption{a) raw flow image b) the filtered flow image.}
	\label{filter_flow}
\end{figure}

All flow images are filtered using the same non-local means filter as used in the segmentation of the pore space. An example of the filtering is shown in Figure \ref{filter_flow}.

\begin{figure}[!htbp]
	\centering
\includegraphics[width=1.0\textwidth]{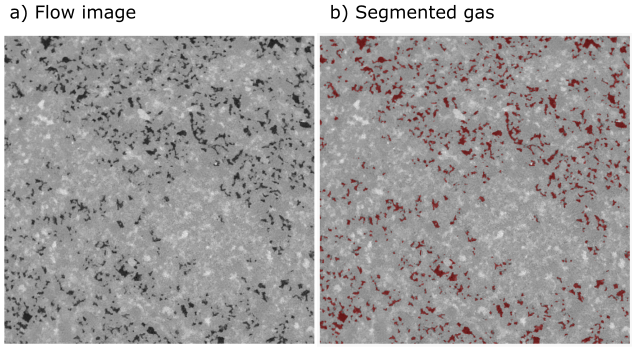} 
	\caption{a) the flow image, and b) the segmented gas (red) overlain on the flow image.}
	\label{gas_segmented}
\end{figure}

To segment the gas, the filtered, registered flow image is subtracted from the filtered, registered brine image. The difference is the location of the gas, which can then be segmented using a simple greyscale threshold. Post processing is carried out by removing small objects (smaller than 4 voxels) and filling in small holes (smaller than 4 voxels). The gas is then masked by the segmented pore space. An example of the gas segmentation is shown in Figure \ref{gas_segmented}b. 

The location of the water is calculated by assigning all locations of the pore space not occupied with gas, as water. This is a valid assumption for two-phase flow. As gas is the non-wetting phase, it occupies the largest pores, and so is the phase of interest in this work.

\section{Results and Discussion}

\subsection{Sensitivity of multiphase flow properties to the image processing workflow}

\begin{figure}[!h]
	\centering
\includegraphics[width=1\textwidth]{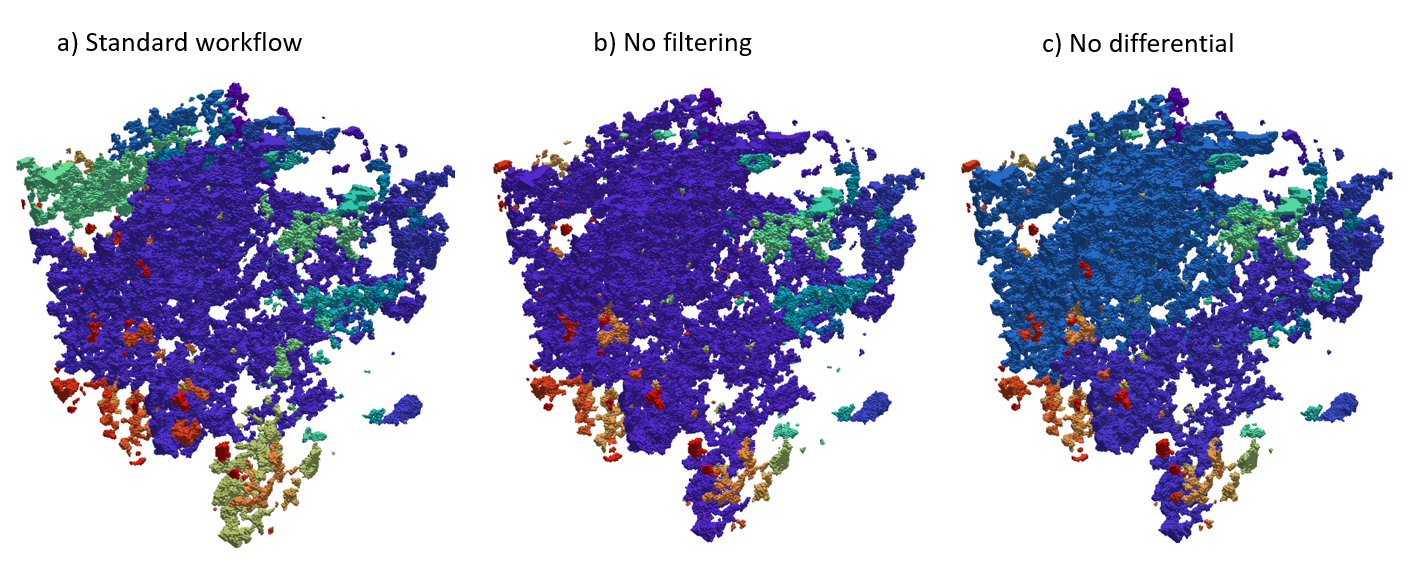} 
	\caption{Gas ganglia for different image processing workflows: a) standard workflow, b) same workflow as a) but with no filtering of the flow images prior to segmentation, c) same workflow as a) but with no differential imaging using the brine scan. Each distinct ganglia is given a different colour to highlight connectivity. The brine and rock grains are transparent.}
	\label{different_workflows}
\end{figure}

\begin{figure}[!h]
	\centering
\includegraphics[width=1\textwidth]{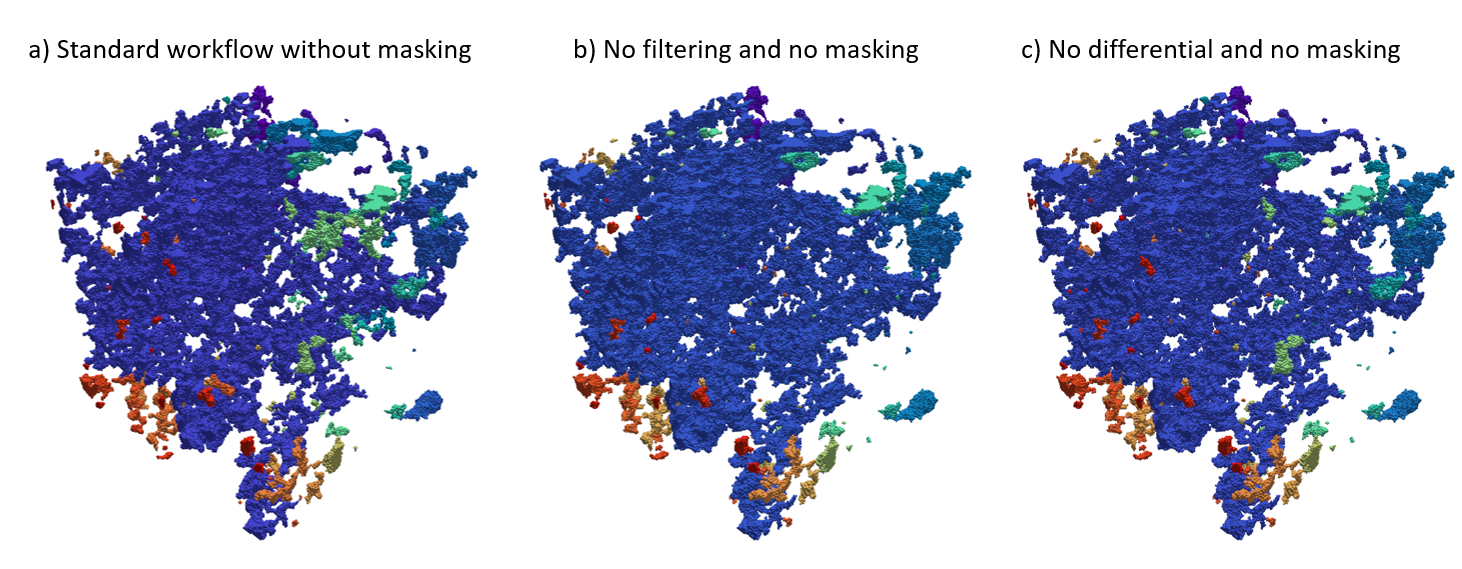} 
	\caption{Gas ganglia for different image processing workflows without masking of the pore space: a) standard workflow, b) same workflow as a) but with no filtering of the flow images prior to segmentation, c) same workflow as a) but with no differential imaging using the brine scan. Each distinct ganglia is given a different colour to highlight connectivity. The brine and rock grains are transparent.}
	\label{different_workflows_no_masking}
\end{figure}

\begin{figure}[!h]
	\centering
\includegraphics[width=1\textwidth]{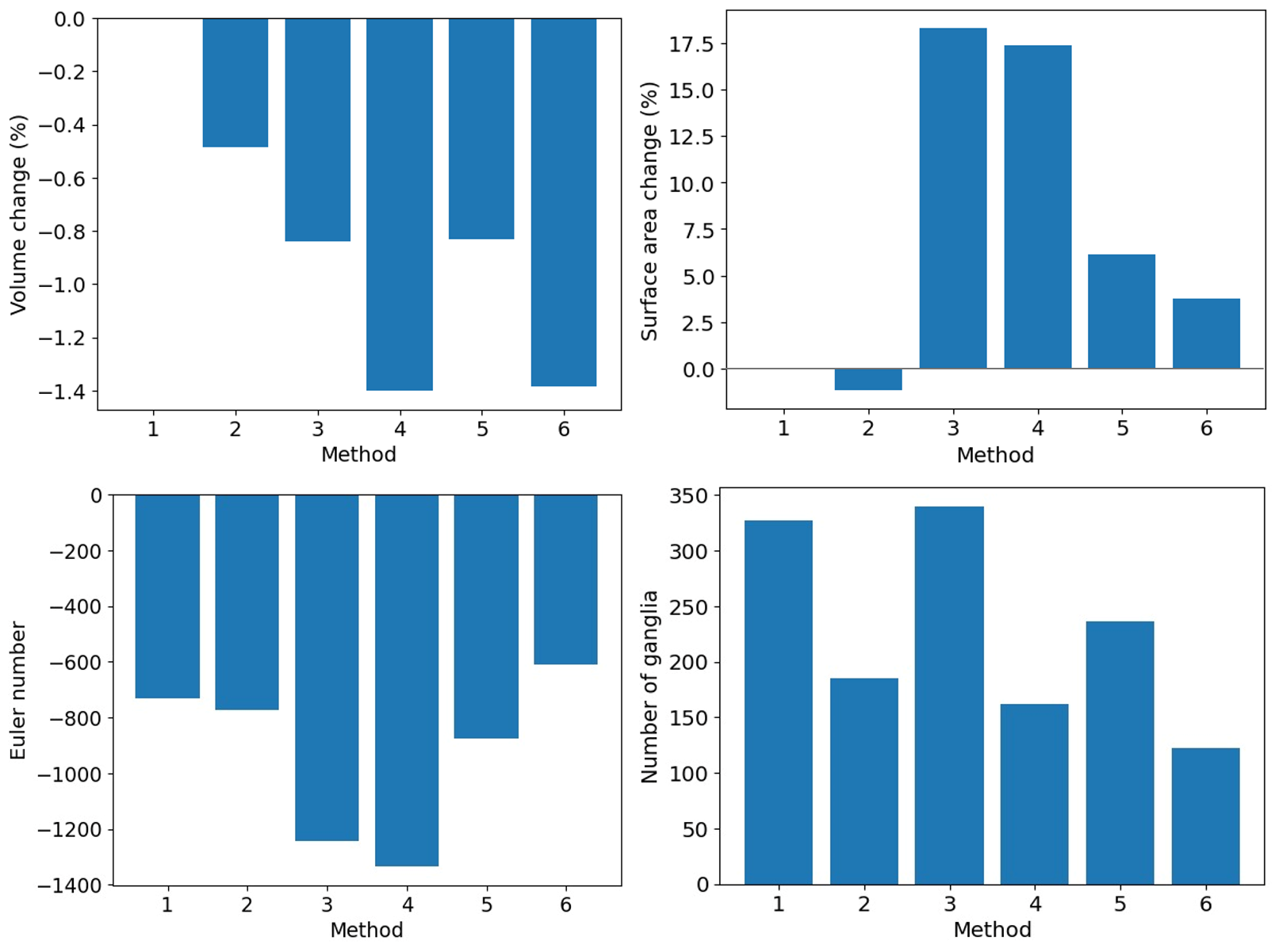} 
	\caption{The impact of image processing workflow on multiphase flow parameters. Method 1 = standard workflow, method 2 = standard workflow without masking, method 3 = no filtering, method 4 = no filtering or masking, method 5 = no differential imaging, and method 6 = no differential imaging or masking.}
	\label{Minkowski_functionals}
\end{figure}

The gas segmentation workflow was varied to understand its impact on the flow image segmentation. The workflow described in Figure \ref{workflow}, is considered the standard workflow. We explored the impact of removing the filtering step and removing the differential image step, with the gas segmentation for the different workflows shown in Figure \ref{different_workflows}. Here, it can qualitatively be observed that the number of gas ganglia is changing. We also explored the impact of removing the masking of the pore space, with this shown in Figure \ref{different_workflows_no_masking}. The impact of these different methods is shown quantitatively in Figure \ref{Minkowski_functionals}.

In Figure \ref{Minkowski_functionals}, it can be observed that the image processing workflow has little affect on the total volume of gas. Masking with the pore space has little impact on volume and surface area. However, the filtering influences the surface area significantly, with no filtering resulting is around 15\% - 18\% more gas surface area. The Euler number for the gas is large and negative regardless of the image processing workflow, but again, removing the filtering step results in a significant change. The masking has little impact on the Euler number, but significantly influences the number of distinct gas ganglia. 

\subsection{Sensitivity of porosity to image resolution}

\begin{figure}[!htbp]
	\centering
\includegraphics[width=1.0\textwidth]{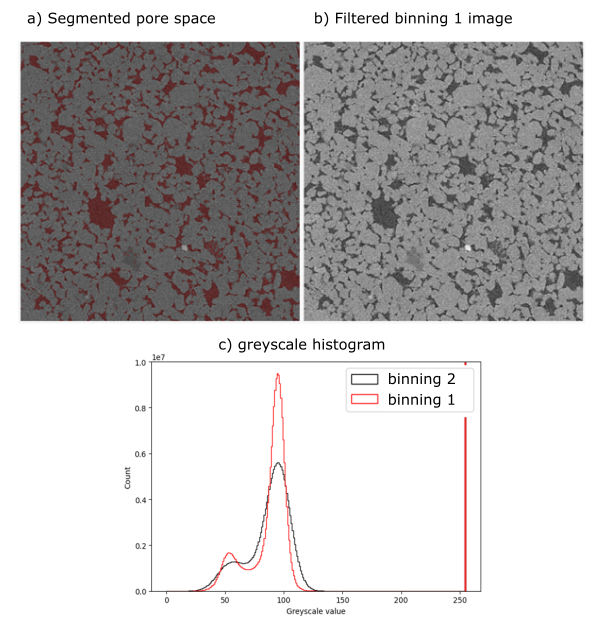} 
	\caption{a) the segmented pore space overlain on the filtered binning 1 image, b) the filtered binning 1 image, and c) the greyscale histograms for the binning 1 and binning 2 images.}
	\label{binning1}
\end{figure}

As observed in Figure \ref{different_workflows} and Figure \ref{different_workflows_no_masking}, the masking of the pore space influences the number of gas ganglia significantly, but has little impact on the volume of gas. There is uncertainty in the segmentation of the pore space due to filtering, segmentation method and image resolution. We explore the role of resolution on the segmentation of the pore space. We have two identical images of the pore space, except in one instance the binning of the image is 1, and in the other the binning is 2. The segmentation of the binning 2 example is shown in Figure \ref{watershed_segmentation}. The segmentation of the binning 1 example is shown in Figure \ref{binning1}. Binning aims to reduce noise, but at the expense of image resolution, so the binning 2 image has a voxel size of 20~$\mu m$ while the binning 1 image has a voxel size of 10~$\mu m$. The segmentation of the binning 1 image takes 6 times longer to run, and the images take up 4 times more memory. Thus the binning 2 images are easier to process, but important information may be lost in the binning process. The increase in resolution in the binning 1 case makes two distinct peaks in the greyscale histogram in Figure \ref{binning1}c), although there is still enough overlap to necessitate use of the watershed segmentation algorithm. 

The porosity for the binning 2 image is 0.138, whereas it is 0.142 for the binning 1 image, a difference of 3\%. Note that a typical Bentheimer sandstone has a porosity between 0.21-0.27 \cite{peksa2015bentheimer}, which implies that around 50\% of the total porosity is below the image resolution. There is a 20\% difference in the connectivity of the pore space, calculated by the number of disconnected regions. These regions are connected via porosity below the resolution of the images, as no region remains filled with gas after brine is injected into the sample, thus all regions of the pore space must be connected to the inlet. This suggests that the increase in resolution picks up smaller connections in the pore space. As the non-wetting phase is more robust to image processing because it occupies the largest pores, it is a more important metric than the pore space connectivity. However, one should be mindful of missing connections when calculating absolute permeability from images, instead of measuring it directly on the sample using a pressure transducer.

\subsection{Sensitivity of connectivity measures to classification of connectivity}
\begin{figure}[!h]
	\centering
\includegraphics[width=1\textwidth]{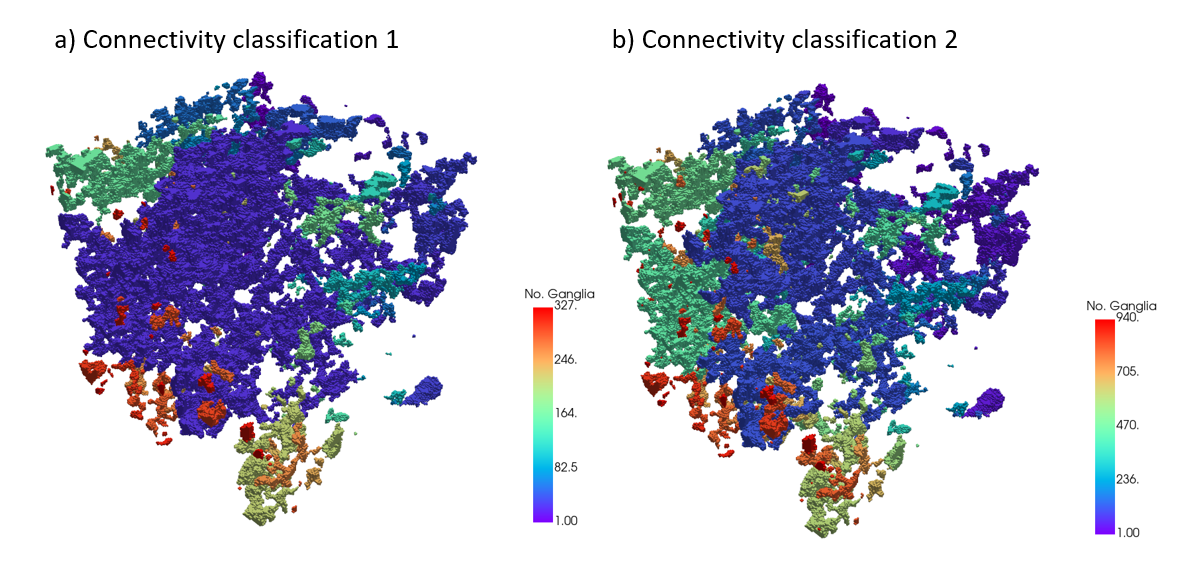} 
	\caption{Gas ganglia distribution for a) the standard workflow where voxels that share a face are considered connected, and b) the standard workflow but voxels that share an edge are considered connected.  }
	\label{different_connectivty_class}
\end{figure}

\begin{figure}[!h]
	\centering
\includegraphics[width=1\textwidth]{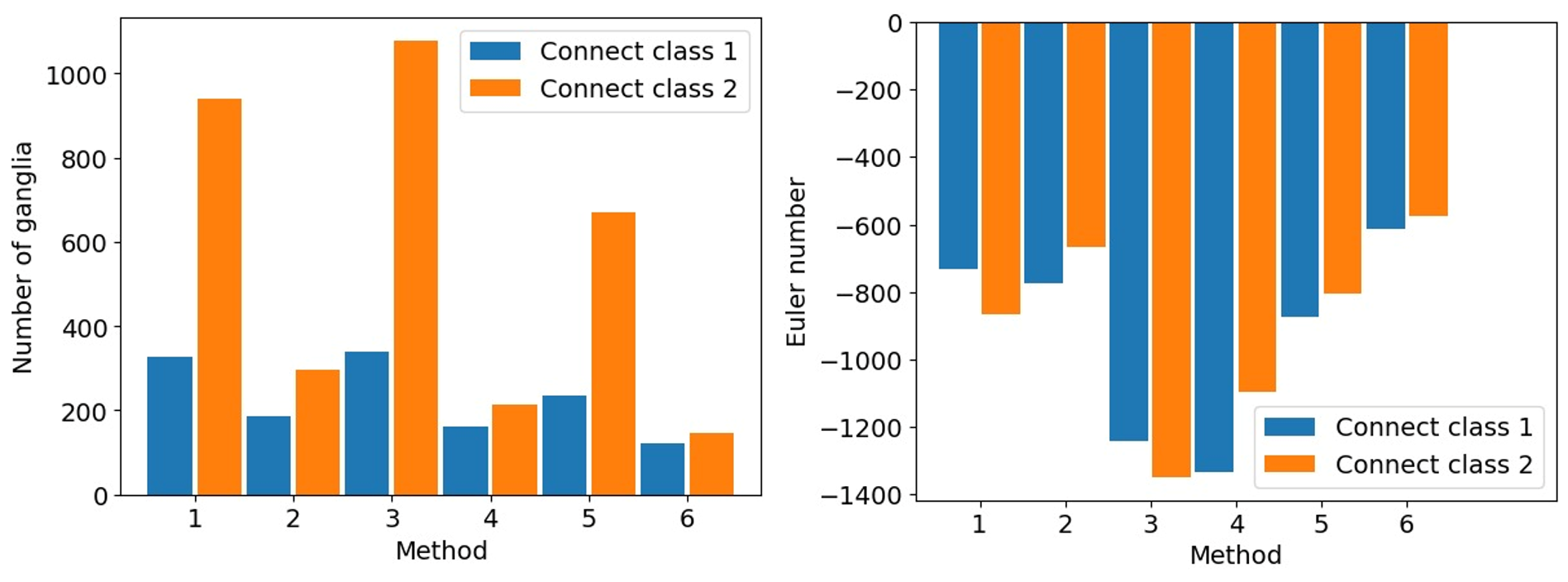} 
	\caption{The impact of connectivity classification on the number of ganglia and Euler number but the different image processing workflows. Connect class 1 implies that voxels must share a face to be considered connected, and connect class 2 classifies voxels with a shared edge as connected. Method 1 = standard workflow, method 2 = standard workflow without masking, method 3 = no filtering, method 4 = no filtering or masking, method 5 = no differential imaging, and method 6 = no differential imaging or masking.}
	\label{connectivity_class_parameters}
\end{figure}

We explore the impact of connectivity classification on the number of ganglia and Euler number. In the standard workflow, we assume that a voxel must share a face with another voxel for these voxels to be considered neighbouring; this is connectivity classification 1 in Figure \ref{different_connectivty_class}. We can update this to assume that a voxel only has to share an edge with another voxel to be considered neighbouring; this is connectivity classification 2 in Figure \ref{different_connectivty_class}. The impact of this classification is shown in Figure \ref{connectivity_class_parameters}.

The number of ganglia is heavily influenced by the connectivity classification, as shown in Figure \ref{connectivity_class_parameters}. The difference between connectivity classifications is smaller without masking of the pore space for all workflows. The impact of connectivity classification is less significant for the Euler number. However, for connectivity classification 2, the connectivity decreases when we have the masking step, and increases when we do not mask the gas with the pore space compared to connectivity classification 1. 

\section{Conclusions}
We present an accurate, consistent, and open-source workflow for processing 3D micro-CT data from multiphase flow experiments. Our workflow is similar in accuracy to other workflows, but can be fully integrated into modelling pipelines, and allows for comparisons between different studies. 

We highlight that multiphase flow properties are sensitive to the image processing workflow. While the total volume of the non-wetting phase is consistent across methods, the surface area and connectivity of the non-wetting phase is significantly influenced by filtering, and the classification of connectivity. By using this workflow, we are able to understand the uncertainty in parameters caused by the image processing workflow.

By establishing a fast and consistent workflow for image processing that accounts for potential variations caused by user bias and methodology, we can provide a large training dataset to augment existing segmentation datasets used to train machine learning models. Although machine learning approaches are more resilient to noisy data, their performance on filtered images is comparable to that of algorithmic workflows \cite{garfi2020sensitivity}. Given the significant influence of filtering on surface area and connectivity, further investigation into these variations is warranted to enhance the accuracy and reliability of the models.

%%%%%%%%%%%%%%%%%%%%%%%%%%%%%%%%%%%%%%%%%%%%%%%

\section{Data availability}
The code used in this manuscript is available on GitHub: 
\url{https://github.com/cspurin/image_processing}. 

\section{CRediT author statement}
\textbf{Catherine Spurin}: Conceptualization, Methodology, Software, Validation, Writing - Original Draft, Writing - Review \& Editing,
\textbf{Sharon Ellman}: Conceptualization, Methodology, Validation, Writing - Original Draft, Writing - Review \& Editing,
\textbf{Dane Sherburn}: Conceptualization, Methodology, Software,
\textbf{Tom Bultreys}: Conceptualization, Supervision,
\textbf{Hamdi Tchelepi}: Supervision.

\section{Acknowledgements}
Catherine Spurin and Hamdi Tchelepi acknowledge support from the GeoCquest consortium. 
Sharon Ellman is a PhD Fellow with the Research Foundation – Flanders (FWO) and acknowledges its support under grant 1182822N. Tom Bultreys holds a senior postdoctoral fellowship from the Research Foundation-Flanders (FWO) under Grant No. 12X0922N. This research also received funding from the Research Foundation–Flanders under grant G051418N, G004820N and the UGent BOF funding for the Centre of Expertise UGCT (BOF.EXP.2017.0007). We also wish to thank Griffin Chure, for his useful tutorials on the python sci-kit library. 

\section{Declaration of Interests}
The authors of this article declare that they have no conflict of interests.

%% ------------------------------------------------------------------------ %%
%% References and Citations

%%%%%%%%%%%%%%%%%%%%%%%%%%%%%%%%%%%%%%%%%%%%%%%
%
% \bibliography{<name of your .bib file>} don't specify the file extension
%
% don't specify bibliographystyle

% In the References section, cite the data/software described in the Availability Statement (this includes primary and processed data used for your research). For details on data/software citation as well as examples, see the Data & Software Citation section of the Data & Software for Authors guidance
% https://www.agu.org/Publish-with-AGU/Publish/Author-Resources/Data-and-Software-for-Authors#citation

%%%%%%%%%%%%%%%%%%%%%%%%%%%%%%%%%%%%%%%%%%%%%%%

\section{Appendix}
\subsection{Registering the flow images to the initial dry image}
The registration of the flow images works by matching features between the two input images. To minimise the computational expense of the registration, we advise performing the registration of the flow images on a smaller subvolume (that captures the entire cross section of the sample) and applying the output to the whole image. The brine saturated image is noisier than the flow images, because the brine is more attenuating than the gas. It is advisable to register the image of a subvolume with a key feature (such as an inclusion), and then apply the registration to the entire volume. 

The axes change after registration, so the image must be resampled to ensure consistency across all images. Resampling involves adjusting the axes to align with the initial image. In our example, the registration involves a simple translation, so resampling merely corrects the starting point of the axes. If shear movement was present in the experiments, then resampling must correct the axes.

\subsection{Merging of sections}
In this work, we explore a subvolume of the dataset. For a machine with 32GB ram, the maximum volume that could be processed using our workflow was $1400 \times 1400 \times 700$ voxels. If the desired image is larger than this, it can be segmented in sections, and merged in post processing. If this is the case, then greyscale values must be consistent when segmenting. In this workflow, the input image is normalised to the maximum and minimum values in the input image prior to segmentation. If the maximum and minimum greyscale value changes for the different sections, this should be overwritten, with the global minimum and maximum for all sections used instead. 

For merging, slices are compared until identical slices are found. If merging the segmented scans, there can be some variation caused by non-local means filtering at the edges. To overcome this, there should be around 100 slices overlap between sections that will be merged. If small movements occur between scans, then merging will fail e.g. a grain moving. Small discrepancies, within the error of the segmentation should be removed. For this work, we remove all objects smaller than 4 voxels in volume. This number can be increased, but should be done with caution.

\bibliography{refs}

\end{document}